\documentclass[prd,aps,twocolumn,nofootinbib,superscriptaddress,
eqsecnum,floatfix,longbibliography]{revtex4-1}

\def\@fpheader{\bigskip\relax}
\makeatother

\usepackage{amsfonts,amssymb,bm}
\usepackage[mathscr]{eucal}
\usepackage{amsmath}
\usepackage{epsfig,graphicx}
\usepackage{bbold}
\usepackage{datetime}
\usepackage[utf8]{inputenc}
\usepackage{ogonek}

\usepackage[dvipsnames]{xcolor}

\def\Rest{Rest }

\newcommand{\be}{\begin{equation}}
\newcommand{\ee}{\end{equation}}
\newcommand{\ba}{\begin{eqnarray}}
\newcommand{\ea}{\end{eqnarray}}
\newcommand{\beg}{\begin{gather*}}
\newcommand{\eng}{\end{gather*}}

\newcommand{\const}{{\rm const}}

\newcommand{\hh}{,\hspace{0.5cm}}

\newcommand{\eq}[1]{(\ref{#1})}

\newcommand{\ins}[1]{{\mbox{\tiny #1}}}

\def\XXint#1#2#3{{\setbox0=\hbox{$#1{#2#3}{\int}$ }
\vcenter{\hbox{$#2#3$ }}\kern-.6\wd0}}

\begin{document}

\title{\boldmath Thomas precession, relativistic torque, and
  non-planar orbits}

\author{Andrzej Czarnecki}
\email{andrzejc@ualberta.ca}
\affiliation{Department of Physics, University of Alberta, Edmonton, Alberta, Canada T6G 2E1}

\author{Andrei Zelnikov}
\email{zelnikov@ualberta.ca}
\affiliation{Department of Physics, University of Alberta, Edmonton, Alberta, Canada T6G 2E1}

\begin{abstract}
We analyze the angular momentum balance for a particle undergoing
Thomas precession. The relationships among relativistic torque, the
center of mass, and the center of inertia for a spinning particle are
clarified. We show that spin precession is accompanied by orbital
angular momentum precession, and present examples of the resulting out-of-plane
motion. 
\end{abstract}
\maketitle

\section{Introduction}

In classical physics, spin is the intrinsic angular momentum of a
rotating body, such as a gyroscope. The spin orientation remains
constant in the absence of external forces, but can change under
acceleration. This relativistic effect, known as Thomas precession
\cite{Thomas:1926dy}, was discovered by
Föppl and Daniell, 
\cite{foppl1913kinematik},
 and independently by
Silberstein \cite{silberstein1914SR}; for further historical context,
see \cite{Krivoruchenko:2008cw}. 

Acceleration also induces radiation. When the spin precesses, does the
orbital angular momentum $\bm L$ adjust to conserve the total angular
momentum $\bm J$, or is the imbalance radiated away? One might also
consider a periodic exchange of angular momentum between the particle
and its near-zone electromagnetic field. 

Here we find that when a charged spinning particle orbits a nucleus,
$\bm{L}$ changes as the spin precesses, causing the particle to move
in and out of its average orbital plane.

We consider the limit of the particle's classical radius much smaller
than the curvature radius of its worldline \cite{Teitelboim:1979px}.
Both radiation and periodic exchange are then negligible in comparison
with the change of $\bm{L}$.

We characterize the resulting out-of-plane motion. This relativistic
effect requires a consistent description of the orbiting particle. The
spin is typically defined in the frame comoving with the particle,
while precession and radiation are more easily described in the frame
centered at the nucleus at rest (Lab). This dual approach has often
led to confusion. We clarify and visualize the relativistic dynamics
of the spinning particle from the perspective of a Lab observer.

Precession of a gyroscope requires torque. However, a force acting on
the gyroscope's center of mass exerts no torque in its rest
frame. Muller \cite{Muller:1991} resolved this apparent paradox, and
R\k{e}bilas \cite{Rebilas:2015} further refined the solution, showing
that while there is no torque in the rest frame, the force
accelerating the gyroscope exerts a torque in the Lab frame. 

To understand the origin of this torque, it is important to
distinguish between the center of mass (CM) and the center of inertia
(CI) \cite{Muller:1991}. A consistent treatment of Thomas precession
requires relativistic generalizations of spin, torque, CM, and
CI. Despite the long history and extensive literature on this topic
(for recent work and further references see \cite{Palge:2023mmq,Schmidt2024}),
we believe that the relativistic description of these quantities still
lacks necessary clarity. In this paper, we outline the peculiarities
of the relativistic motion of a spinning particle.

We assume the nucleus is pointlike and sufficiently massive that its 
motion and magnetic moment are negligible. The central electric force 
it exerts on the particle is the only force we consider, which we 
refer to as the force. We assume that the electric field 
does not probe the particle's internal structure. 

We refer to an instantaneous inertial frame comoving with the particle
as the \Rest frame and mark with a bar  quantities defined
in that frame. The Lab frame time is denoted by $t$, and the
particle's proper time by $\tau$. 

Section~\ref{CMCI} clarifies the differences between CM and CI. In
Section~\ref{BMT}, we introduce the kinematics and derive the
evolution of the spin with respect to $\tau$, known as Fermi
transport. Section~\ref{Sec4} discusses relativistic angular momentum
and spin.
 In Section~\ref{Section6}, we
present examples of spin precession and illustrate its dynamics with
plots. Additional examples and technical details related to the
dynamics of CM and CI are provided in Appendices~\ref{AppendixA} and
\ref{AppendixB}.

We use the metric signature $(-,+,+,+)$  and  adopt units with the
speed of light $c = 1$. Summation is implied over
repeated indices. Greek letters ($\alpha, \beta, \dots = 0,1,2,3$)
denote spacetime components, while Latin letters ($i, j, \dots =
1,2,3$) denote spatial components. We work in Cartesian coordinates,
and for the Levi-Civita symbol, we adopt the sign convention
$\varepsilon_{0123} = 1 = -\varepsilon^{0123}$, with
$\varepsilon_{123} = \varepsilon^{123} = 1$ \cite{Misner:1974qy}. 

\section{Centers of inertia and of mass \label{CMCI}}

In this section, we carefully define the centers of inertia and mass
and provide an intuitive explanation of their differences.

The particle's inertial properties are described by the stress-energy
tensor $T^{\alpha\beta}$,  obtained by varying the classical action
with respect to the metric. This tensor is symmetric,
$T^{\alpha\beta} = T^{\beta\alpha}$ \cite{Landau:1975pou}. We use it
to study the particle’s dynamics in the Lab frame. The energy of the
system is the time component $P^0$ of the momentum four-vector
$P^\alpha$,
\be\label{P0} 
P^{0}(t)=\int {\mathrm d}^3x \,T^{0 0}(x).
\ee
 The integral is computed over a hypersurface of events that are
simultaneous in the Lab frame, with $t = \const$.

The position of the particle's center of inertia (CI;  also called ``Lab frame
centroid" \cite{Gralla:2010xg}) in the Lab frame,
\be\label{CI}
R^{\alpha}_\ins{CI}(t)={1\over P^0}
  \int {\mathrm d}^3x \,x^{\alpha} T^{0 0}(x),
\ee
corresponds to our intuition about where the
total energy of the system is centered. The numerator in
Eq.~\eqref{CI}  is the dipole moment of the energy of the system.
Concentrating the whole energy of the system in its
CI does not change its energy dipole moment \cite{Lorce:2018zpf}.

Unlike CI, the position of the  CM,
\be\label{barR}
\bar{R}^{\alpha}= \left(\bar{t},
    {\int {\mathrm d}^3\bar{x}~\bar{x}^i\bar{T}^{0 0}(\bar{x})\over
     \int {\mathrm d}^3\bar{x}~\bar{T}^{0 0}(\bar{x})}
 \right),
\ee
 is not Lorentz
covariant: it depends on the hypersurface $\bar{t}=\const$ in the integrals. Although CM and CI
coincide in the \Rest frame, they may differ in the Lab frame.

The difference between CI and CM can be understood with the example of
a bicycle wheel \cite{Muller:1991}. In the frame of a street, the CI
of a moving wheel is above the hub because the speeds of the spokes
are larger there than below the hub (the point of the wheel-street
contact is at rest with respect to the street, while the top of the
wheel is moving with twice the speed of the bike). Elements of the
wheel above the hub have more kinetic energy.

To explain the interplay of the dynamics of the particle's spin and
its orbital motion, we define the part of the total angular momentum
which describes the intrinsic rotation of the particle. The orbit is
the trajectory of the CM. It is moving with the four-velocity
$U^{\alpha}=P^{\alpha}/M$ \cite{Lorce:2018zpf}, where
$M=\sqrt{-P^{\alpha}P_{\alpha}}$ is the particle's mass. This
definition formalizes the intuition that the particle is at rest in
the CM rest frame and the Lorentz invariant energy
$\bar{P}^{0}=-P^{\alpha}U_{\alpha}$ equals the invariant mass
$M$. From now on we use this choice of $U^{\alpha}$ to describe the CM
velocity.

For an isolated system the total stress-energy tensor is conserved,
$\partial_\lambda T^{\alpha\lambda}=0$,
hence, in the absence of radiation to infinity, momentum
$P^{\alpha}$ is constant. If an external force, such as an external
Maxwell field, acts on the particle, then
$
\partial_\lambda T^{\alpha\lambda}=f^{\alpha},
$
where the four-vector $f^{\alpha}(x)$ is the density of the external force.

\section{Fermi transport \label{BMT}}

Fermi transport will be a key tool in our discussion of the orbital
motion of a spinning particle. The crucial result of this section is
Eq.~\eqref{Fermi}, which describes the evolution of the spin with
respect to proper time. 

In classical physics, we consider the spin as a three-vector
$\mathbf{S}$. In special relativity, a vector that is purely spatial
in the \Rest frame acquires a time component in
the Lab frame. Thus, we treat the spin as a four-vector.

In the Lab frame, the particle's worldline has coordinates
$R^{\alpha}(t)=\big(R^0(t),R^{i}(t)\big)$ 
with $R^0(t)=t$.
The four-velocity $U^{\alpha}$ of the particle is
$U^{\alpha}={d R^{\alpha}\over d\tau}$.
The worldline of a massive particle is timelike,
\be\label{UU}
U^{\alpha}U_{\alpha}=-1.
\ee
In the \Rest frame, $\bar{U}^{\alpha} = (1, 0, 0, 0)$. The fact that
the spin four-vector $S^{\alpha}$ reduces to a spatial vector in the
\Rest frame can be invariantly expressed through orthogonality,
\be\label{SU}
S^{\alpha}U_{\alpha}=0.
\ee
Spin's magnitude is constant,
\be\label{SS}
S^{\alpha}S_{\alpha}=\const.
\ee
In the Lab frame the four-velocity reads
$U=\gamma (1,\bm V)$ with $\bm V={d \bm R/dt}$ and
$\gamma=1/\sqrt{1-\bm V^2}$.
Because of \eq{UU}, four-acceleration, $w^{\alpha}={d U^{\alpha}\over d\tau}$, is  orthogonal to the four-velocity
\be\label{Uw}
w^{\alpha}U_{\alpha}=0.
\ee
Orthogonality \eq{SU} and  constant magnitude
\eq{SS} have to be satisfied at any point of the worldline,
\be\label{dSUdSS}
{d\over d\tau}(S^{\alpha}U_{\alpha})=0 \hh {d\over d\tau}(S^{\alpha}S_{\alpha})=0.
\ee
The four-dimensional rotation, or infinitesimal Lorentz
transformation, is defined by an antisymmetric matrix, 
\be
{dS^{\alpha}\over d\tau}=\Omega^{\alpha\beta}S_{\beta} \hh
 \Omega^{\alpha\beta}=-\Omega^{\beta\alpha}.
\ee
$\Omega^{\alpha\beta}S_{\alpha}S_{\beta}=0$ ensures the second condition \eq{dSUdSS}.

The matrix $\Omega^{\alpha\beta}$ has non-zero components only in the
plane of rotation.  It is useful to think about rotation in a
plane (it can be defined in arbitrary dimensions greater
than one) rather than around an axis which can be defined only in
three dimensions \cite{Misner:1974qy}.

The rotation plane is defined by the four-velocity $U^{\alpha}$ and
the four-acceleration $w^{\alpha}$. The only antisymmetric structure
defined by these vectors has the form \cite{Misner:1974qy}
\be\label{FW}
\Omega^{\alpha\beta}=a(U^{\alpha}w^{\beta}-w^{\alpha}U^{\beta}).
\ee
$a=1$ follows from the first condition in
\eq{dSUdSS}.
Taking into account conditions \eq{SU} and \eq{Uw}, we obtain
\be\label{Fermi}
{dS^{\alpha}\over d\tau}=U^{\alpha}w^{\beta}S_{\beta}.
\ee
This equation defines a  transport law of  the spin vector along an arbitrary worldline.
Proposed by Fermi \cite{Fermi1922}, it is known as the Fermi
transport.
It leads to the Bargmann–Michel–Telegdi  (BMT) equation \cite{Bargmann:1959gz}
for the dynamics of the spin
(see \cite{Rebilas:2011} for a simple derivation).

\section{Orbital and spin angular momenta}\label{Sec4}
We now derive the crucial Eq.~\eqref{DeltaR} for the difference of
positions of the centers of inertia and of mass.

Consider the
orbital angular momentum $L^{\alpha}$, the spin $S^{\alpha}$, and
the total angular momentum $J^{\alpha}=L^{\alpha}+S^{\alpha}$. Instead
of  these vectors it is useful to introduce their dual
antisymmetric tensors
$J^{\alpha\beta}$, $L^{\alpha\beta}$, and $S^{\alpha\beta}$, defined
by  (similarly for $J$ and $L$)
\be\label{Lvec2}
S_{\alpha}=-\frac{1}{2} \varepsilon_{\alpha\rho\mu\nu}U^{\rho}S^{\mu\nu}, \hskip 0.8cm
S^{\mu\nu}=\varepsilon^{\mu\nu\alpha\beta}U_{\alpha}S_{\beta}.
\ee
$J^{\alpha\beta}$
is given (Ref.~\cite{Misner:1974qy}, \S 5.11) by the integral over a
spacelike surface $t=\const$ in Lab,
\be\label{Jab1}
J^{\alpha\beta}=\int {\mathrm d}^3x \,\Big[
x^{\alpha}T^{0\beta}-x^{\beta}T^{0\alpha}\Big],
\ee
The force being central, $J^{\alpha\beta}$ is conserved, provided that we neglect
back-reaction effects of radiation to infinity.
The orbital angular momentum is
\be\begin{split}\label{Lab2}
L^{\alpha\beta}(t)&=\int {\mathrm d}^3x \,\Big[
R^{\alpha}(t)T^{0\beta}-R^{\beta}(t)T^{0\alpha}\Big]\\
&=R^{\alpha}(t)P^{\beta}-R^{\beta}(t)P^{\alpha},
\end{split}\ee
where $R^{\alpha}(t)$ is the worldline of CM. Both
orbital $L^{\alpha\beta}(t)$ and intrinsic $S^{\alpha\beta}(t)$
angular momenta depend on $t$.

Similarly to \eq{SU} the spin tensor is orthogonal to the
four-velocity (the supplementary spin condition \cite{Costa:2014nta}),
\be\label{SabUb} 
S^{\alpha\beta}U_{\beta}=0,
\ee
Fermi transport \eq{Fermi} for the spin tensor is
\be\label{dSmn}
{dS^{\alpha\beta}\over d\tau}=-\big(U^{\alpha}S^{\beta\lambda}-U^{\beta}S^{\alpha\lambda}\big) w_{\lambda}.
\ee
Using definitions \eq{Jab1}, \eq{Lab2} one writes
\be\label{SabTa}
S^{\alpha\beta}(t)=\int {\mathrm d}^3x \,\Big[
\big(x^{\alpha}-R^{\alpha}(t)\big)T^{0\beta}-\big(x^{\beta}
-R^{\beta}(t)\big)T^{0\alpha}
\Big].
\ee
In Lab, $R^{\alpha}(t)=x^0=t$, and $S^{\alpha0}$ becomes
\be\label{Sa0}
S^{\alpha0}(t)=\int {\mathrm d}^3x \,
\big(x^{\alpha}-R^{\alpha}(t)\big)T^{00},
\ee
the energy dipole moment about CM.
Using the CI definition \eq{CI} and the total energy  \eq{P0},
\be\label{Sa00}
S^{\alpha 0}(t)=P^0\big[
R^{\alpha}_\ins{CI}(t)-R^{\alpha}(t)\big].
\ee
The shift of the CI relative to the CM is determined by the energy and
the $S^{\alpha 0}$ spin tensor component,
\be\label{DeltaR}
\Delta R^{\alpha}(t)=R^{\alpha}_\ins{CI}(t)-R^{\alpha}(t)={S^{\alpha 0}(t)\over
P^0}.
\ee
This shift of the CI is zero in the \Rest frame but not in the Lab frame.

\section{Relativistic spin precession}\label{Section6}

We consider a particle moving along a given trajectory. For simplicity
we choose the motion of CM to be a circle and study the dynamics of
the spin and its relation to $\Delta R^{\alpha}$.
The back-reaction of the spin on the motion of CM is considered in
Appendix \ref{AppendixB}.

Let $r$ be a radius of the orbit and $\omega$ the frequency of
rotation along the orbit. Then
\be
R^{\alpha}=(t,R^x,R^y,R^z)=(t,r\sin\omega t,r\cos\omega t, 0).
\ee
\begin{figure}[tbp]
\centering
\includegraphics[scale=0.7]{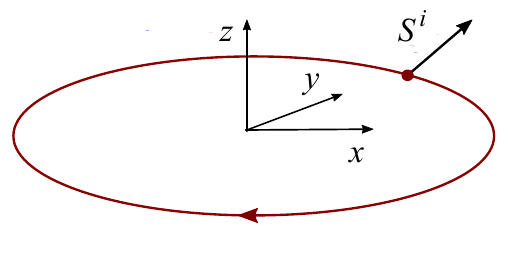}
  \caption{The worldline $R^i(t)$ of the particle with the spin $S^i$.
\label{TP-Figure-1}}
\end{figure}
The four-velocity and four-acceleration read
\begin{align}
U^{\alpha}&={dR^\alpha\over d\tau}=\gamma~ (1, V^x,V^y,V^z)\\
&=\gamma~ (1, V\,\cos\omega t, -V\,\sin\omega t,0), \\
w^\alpha & ={dU^\alpha\over d\tau}
=-\gamma^2 \omega^2 r ~[0,\,  \sin\omega t,\,\cos \omega t,\,0].
\label{wmu}
\end{align}
Fermi transport  \eq{Fermi} leads to $\partial_t S^z=0$ and
\be\begin{split}
&\partial_t S^0=-\gamma^2\omega^2 r\,(\sin\omega t ~S^x+\cos\omega t~S^y),\\
&\partial_t S^x=-\gamma^2\omega^3 r^2 \cos\omega t~(\sin\omega t ~S^x+\cos\omega t~S^y),\\
&\partial_t S^y=+\gamma^2\omega^3 r^2 \sin\omega t~(\sin\omega t ~S^x+\cos\omega t~S^y).
\end{split}\ee
The solution (p.~175 in Ref.~\cite{Misner:1974qy}) is $S^z=\const $ and
\begin{align}\label{Sxy}
S^0&=\tilde{S}\,\sqrt{\gamma^2-1}\cos\gamma\omega t,\nonumber\\
S^x&=\tilde{S}\Big({\gamma+1\over 2}\cos[(\gamma-1)\omega t]+{\gamma-1\over 2}\cos[(\gamma+1)\omega t]\Big),\nonumber\\
S^y&=\tilde{S}\Big({\gamma+1\over 2}\sin[(\gamma-1)\omega
     t]-{\gamma-1\over 2}\sin[(\gamma+1)\omega t]\Big).
\end{align}
The Thomas precession rate $\Omega_T$ can  be read off the first terms
in \eq{Sxy},
$\Omega_T=(\gamma-1)\omega.$
The direction of rotation is opposite to the rotation of the particle,
as can be deduced from comparison with the direction of rotation of
the velocity vector
$V^{x,y}=V (\cos\omega t,- \sin\omega t)$.
The second terms in Eqs.~\eq{Sxy} describe
oscillations  with frequency $(\gamma+1)\omega$.
Their magnitude,  proportional to
$(\gamma-1)$, is small in the non-relativistic case
when $\gamma-1\to V^2/2\ll 1$.

 Eq.~\eq{Lvec2} provides $(0 i)$ components of the spin  tensor,
\begin{align}
\label{Sxzyz}
&\left[S^{0x}, S^{0y}, S^{0z} \right] = \sqrt{\gamma^2-1}
\nonumber \\ &\qquad
\cdot
\left[ S^z \sin\omega t, S^z \cos\omega t, -\tilde{S}\sin\gamma\omega t \right]. 
\end{align}
Substituted into  \eq{DeltaR},  these solutions give the CI shift in
terms of the invariant mass $M=P^0/\gamma$,
\begin{align}
\label{DeltaR1}
&\left[ \Delta R^{x}(t), \Delta R^y(t), \Delta R^z(t)  \right]
={ V\over M}
 \nonumber \\ &\qquad \cdot
\left[ -S^z \sin\omega t, -S^z \cos\omega t, \tilde{S} \sin\gamma\omega t \right]. 
\end{align}
This result is valid for any value of $\gamma$.
Definition \eq{DeltaR1} of $\Delta R^{\alpha}$ differs from
Eqs.~(16--18) in Ref.~\cite{Rebilas:2015}  by a factor
$\gamma$, inconsequential for their non-relativistic analysis.

Eqs.~\eq{DeltaR1} show that the
dynamics of CI is a superposition of two motions with
frequencies $\omega$ and $\gamma\omega$. Thomas precession
frequency $\Omega_T$  equals their difference while the frequency
of the oscillations is their sum. In the plane of the orbit CI
is moving synchronously with the position of the particle. Projection
of the CI orbit on the $xy$ plane moves in a circle with 
radius $R_\ins{CI}=r - S^z V/ M$ with frequency
$\omega$. Depending on the sign of $S^z$ this circle can be either
inside or outside the CM orbit. $\Delta R^{z}$
describes the shift of CI above and below the orbital plane with
frequency $\gamma\omega$.
For an inertial motion the relativistic shift of the
CI was defined in Eq.~(48) of Ref.~\cite{Lorce:2018zpf}.

\begin{figure}[tbph]
\centering
\includegraphics[scale=0.3]{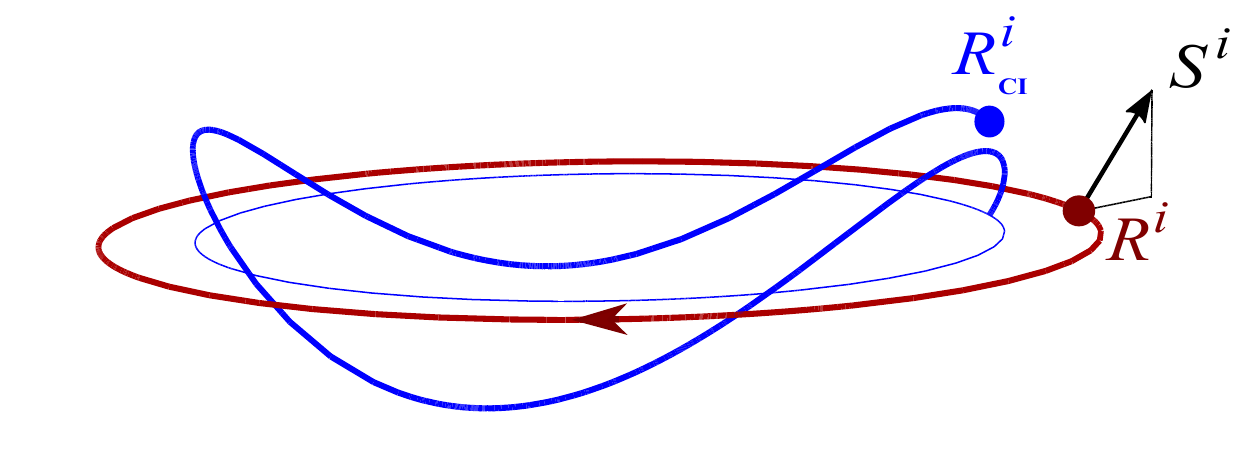}
  \caption{Typical trajectory of the CI (blue curve), when
    the particle moves in a circle. Here the spin's tilt angle is $\pi/4$, ${\bar{S}/ M}=0.3$, and $\gamma=2.3$. 
Projection of the CI trajectory  onto
    the orbital plane is the thin
    (blue) circle  with radius $r-VS^z/M$.
\label{TP-Figure-2}}
\end{figure}

\begin{figure}[tbph]
\centering
\includegraphics[scale=0.3]{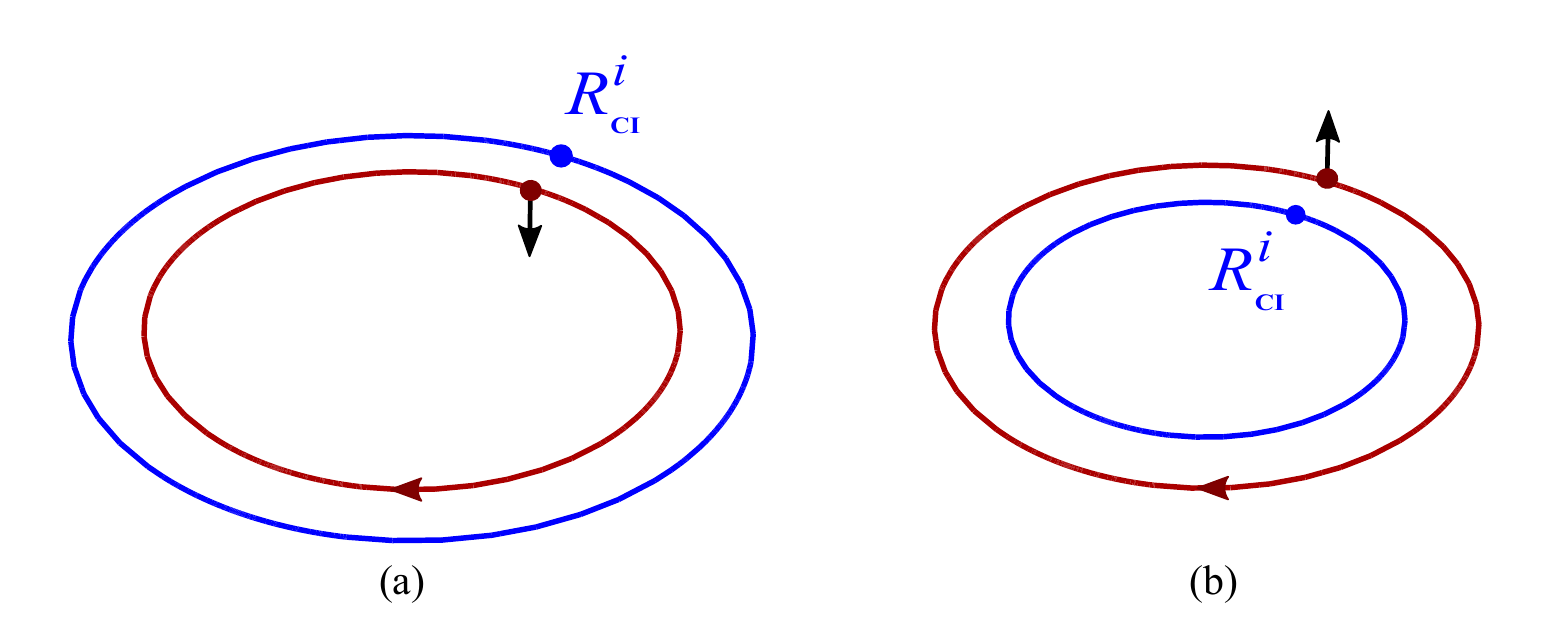}
  \caption{(a) When the spin is parallel to the
    orbital angular momentum $L^i$,  the CI orbit has a
    larger radius than that of CM. (b) If the spin is antiparallel to
    $L^i$, CI is moving in a circle with a smaller radius.
\label{TP-Figure-3}}
\end{figure}

\begin{figure}[tbph]
\centering
\includegraphics[scale=0.6]{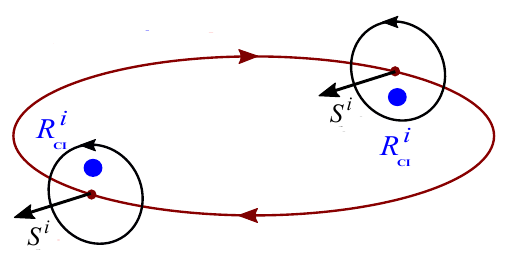}
  \caption{When the spin lies in the orbital
    plane, the CI shifts in the $z$ direction above or
    below the orbital plane depending on the orientation of the spin
    with respect to the velocity.
\label{TP-Figure-4}}
\end{figure}

\begin{figure}[tbph]
\centering
\includegraphics[scale=0.25]{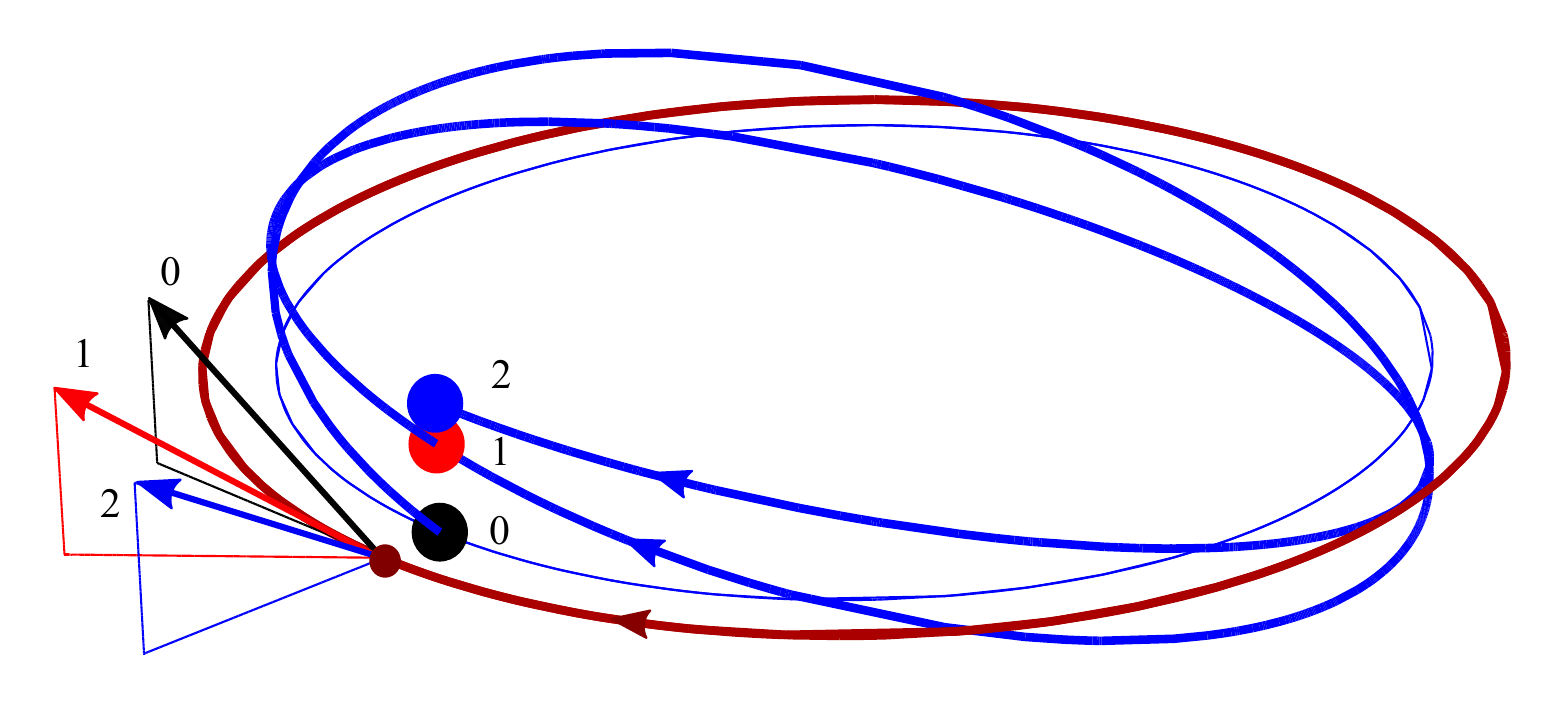}
  \caption{For small velocities the CI orbit is approximately a
    plane   slightly tilted relative to the CM orbit
    and its orientation precesses with the Thomas
    frequency. Three spin vectors  represent orientations of
    the spin at $t=0$ and after two turns
    around the CM orbit ($t=2,4\pi/\omega$). Thin
    (blue) circle is the projection of the CI trajectory onto
    the orbital plane.
\label{TP-Figure-5}}
\end{figure}

\begin{figure}[tbph]
\centering
\includegraphics[scale=0.21]{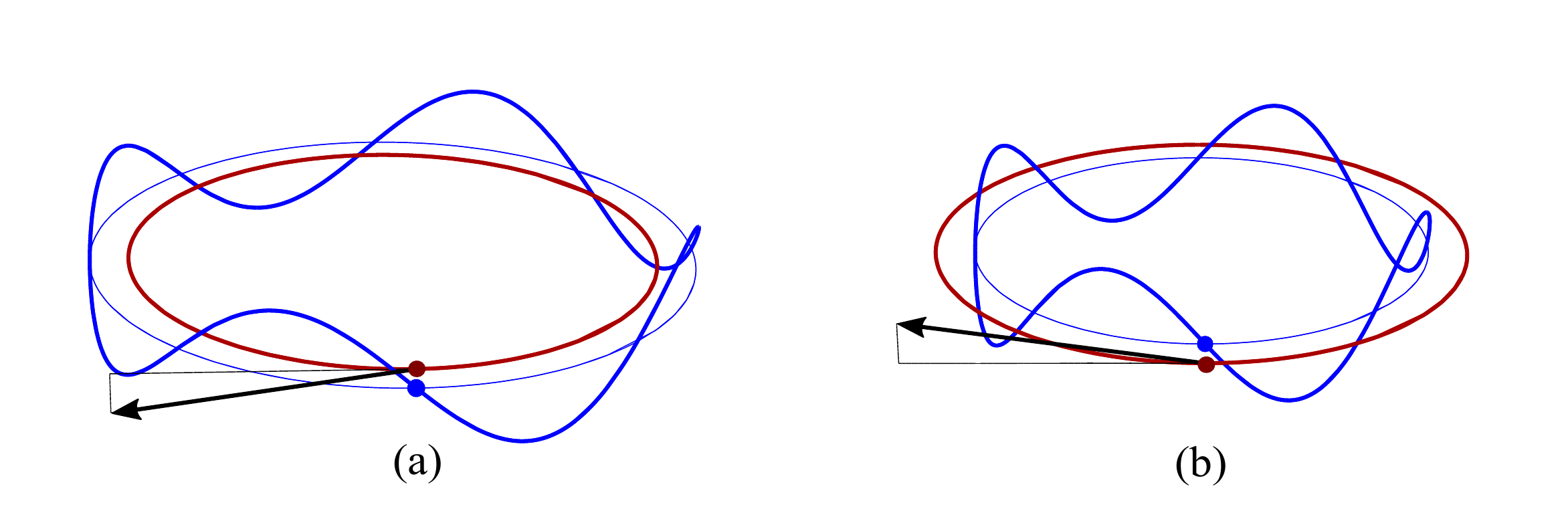}
  \caption{When $\gamma$ is rational, the CI trajectory is
    closed. Here  two orientations of the spin are shown for $\gamma=5$.
\label{TP-Figure-6}}
\end{figure}

Figs.~\ref{TP-Figure-2}-\ref{TP-Figure-6} show CI
trajectories for various spin magnitudes, orientations, and
$\gamma$ factors. For comparable   spin and orbital angular momenta,  CI can move even along
a line orthogonal to the CM orbital plane.

A generic trajectory is illustrated in Fig.~\ref{TP-Figure-2}. Here
the orbital angular momentum is pointing downwards
(negative) while the spin has a positive $S^z$ component. This is why
CI moves inside the circular CM orbit. The projection of the CI
trajectory  is the thin (blue) line. It is a circle. Because
the spin also has non-vanishing components in the orbital plane, the
CI trajectory  wiggles above and below the orbital plane
with frequency $\gamma\omega$.

If the spin $S^i$ is orthogonal to the orbital plane, CI moves along
the circle inside or outside the CM orbit, depending on the sign of
the $S^z$ component, see Fig.~\ref{TP-Figure-3}.

The direction of the CI shift can be understood with a model of an
extended rotating body. For parallel spin and orbital angular momenta,
$\Delta R^{i}(t)$ is in the orbital plane. In this case the velocities
of the orbital motion of the CM and of the outer parts of the rotating
body add up, while in the inner part they partially cancel. Therefore
the relativistic ${\gamma}$ factor is larger in the outer parts.  The
shift of the CI is the difference $x^{\alpha}-R^{\alpha}(t)$ averaged
with the weight $T^{00}(x)/P^{0}$. The ${\gamma}$ factor in
$T^{00}$ gives outer parts a larger weight and the CI shifts
radially outwards.

For the spin antiparallel to the orbital angular momentum, the
same logic dictates that the CI shifts
radially inwards. In both cases the CI trajectory
is circular. If the spin is considerably larger than the orbital
momentum,  $\Delta R^{i}$ can exceed the radius of the
orbit. For an antiparallel spin, the shift
overshoots the orbital center and the
CI is on the opposite
side of the nucleus. The direction and the frequency of rotation
coincide with those of the orbital motion. For a large
antiparallel spin the CI can shift even beyond the CM orbit.

If the spin $S^i$ lies in the orbital plane, the CI shift is
orthogonal to the plane (see Fig.~\ref{TP-Figure-4}). The CI
trajectory lies on a cylinder of the same radius as the CM orbit. It
oscillates above and below the orbital plane with the frequency
$\gamma\omega$, with $\gamma$ accounting for the Thomas
precession. Without precession the spin would always point in the same
direction and the frequency of oscillations about the orbital plane
would be $\omega$. The difference of oscillation frequencies of
$R^{x,y}_\ins{CI}$ and $R^{z}_\ins{CI}$ components is the Thomas
frequency $(\gamma-1)\omega$ as measured in the Lab frame.

The direction of the CI shift is shown in Fig.~\ref{TP-Figure-4}. If
the upper parts of the spinning matter move in the same direction as
the orbital motion, the shift is up. Otherwise it is down. When the
spin is parallel to the orbital velocity, relativistic
factors above and below the plane are the same and the shift
vanishes. Detailed computations of CI for this gyroscope are shown in
Appendix \ref{AppendixA}.

For a generic orientation of the spin,  the CI motion  is a linear
combination of the previous two
cases. The trajectory lies on a cylinder with the radius determined by
the $z$ component of the spin, as in the case of an
(anti)parallel  spin. Seen from the pole, the trajectory is circular. 
From an arbitrary observation point,
Fig.~\ref{TP-Figure-2}, we see a linear combination of an orbital
motion with the frequency $\omega$ and oscillatory motions above and
below the equatorial plane with frequency $\gamma\omega$.

For a non-relativistic orbital motion, $\gamma-1\simeq
V^2/2\ll 1$, CI trajectory resembles a circle in a slightly
tilted plane whose orientation slowly evolves with the Thomas
frequency $\omega V^2/2$. Spin rotates, as depicted in
Fig~\ref{TP-Figure-5}.

\begin{figure}[tbph]
\centering
\includegraphics[scale=0.35]{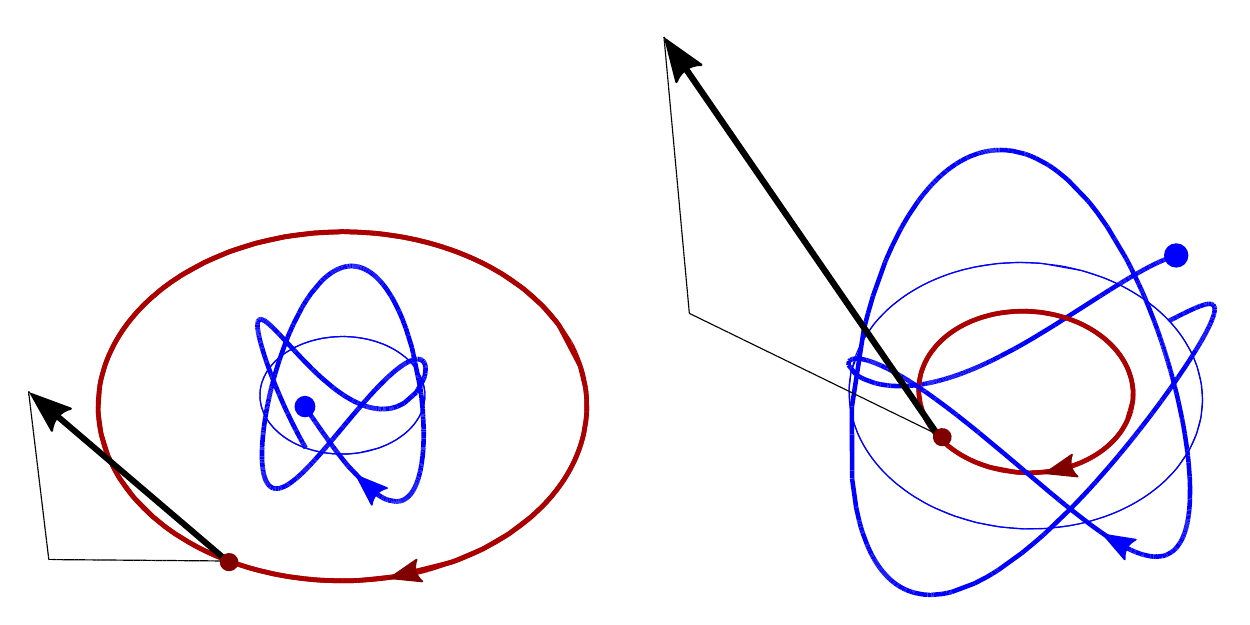}
  \caption{If the spin of the particle has a large $z$ component of
    opposite sign than the orbital momentum, the CI can  shift close
    to the orbital center. For larger spins CI may shift to the other side of the orbit.
\label{TP-Figure-7}}
\end{figure}

Figs.~\ref{TP-Figure-6} and \ref{TP-Figure-7}  present
the CI motion for various orientations
of the spin and velocity. If $\gamma$ is rational, out of plane
oscillations and the orbital rotation 
synchronize and CI moves along a closed curve, see Fig.~\ref{TP-Figure-6}.

In Fig.~\ref{TP-Figure-7} magnitudes of
the spin and the orbital momentum are comparable. Remarkably, the
CI can be on the same side of the nucleus as CM  or, for very
large spins, on the other side. As its radius shrinks, the cylinder on which the CI trajectory is
winding can degenerate 
to a line along which the CI
oscillates  with frequency $\gamma\omega$.

As we have seen, the distance between CM and CI can evolve in a
sophisticated way. Spin precesses but the total angular momentum
remains constant. Naturally, spin should affect the orbital angular
momentum and, as a result, the motion of CM. The
dynamics of the linear momentum of a spinning charged particle was
solved by Frenkel \cite{Frenkel:1926zz}. Contribution of
radiation was derived later \cite{Teitelboim:1979px}. Appendix
\ref{AppendixB} describes this interesting effect.

\section{Discussion}

We have described Thomas precession of a spinning point particle,
assuming that its center of mass (CM) moves along a given trajectory
and that energy is conserved. At any moment, this trajectory can be
considered as a segment of a circular orbit. Since our analysis is
local in time, our conclusions apply not only to circular orbits but
also to more complicated trajectories, provided that energy is
conserved. 

However, if the accelerated particle is charged, it radiates
energy. Fortunately, Thomas precession appears in the leading (first)
order of acceleration, while radiation effects are of the second order
in both acceleration and charge $q$. The effect of radiation on Thomas
precession is negligible, provided that the self-interaction terms
(which are proportional to $q^2$ and hence to the square of the
acceleration) in the equation of motion are small compared to the
external force (which is proportional to $q$). This assumption applies
to a point particle, whose structure is not influenced by the external
field, and where the radius of curvature of the worldline is large
compared to the classical radius $\sim q^2/m$. This argument accounts for
the electromagnetic energy of the charge and applies equally to
spinless particles. Similarly, periodic exchange of angular momentum
with the  particle's near-zone electromagnetic field is a higher-order
effect in $q$ and thus negligible in our approximation. 

In the case of spinning particles, another effect arises: the back
reaction of the spin on the center of mass (CM) orbit of the
particle. This effect is described by the Frenkel equation
(\ref{Frenkel}) \cite{Frenkel:1926zz,Corben:1961zz}, as detailed in
Appendix \ref{AppendixB}. 

The back-reaction correction to the orbit is also of the second order
in acceleration. Thus, similar to the radiation effects, it is small
in the point particle approximation when the spin is smaller than the
orbital angular momentum. 

In Fig.~\ref{TP-Figure-7}, we chose large spin values only for
visualization purposes. Corrections to the point particle orbit due to
spin and radiation-related self-interaction should be considered
perturbatively. 

Due to the spin-orbit interaction, the motion of a spinning particle
can be quite complicated. Such complex behavior has also been observed
in gravitational physics. The motion of a spinning particle in the
gravitational field of a black hole is described by the
Mathisson–Papapetrou–Dixon equations
\cite{Mathisson:1937zz,Mathisson:2010ui,Papapetrou:1951pa,Dixon:1970zza},
which serve as the gravitational analogue of Eq.~\eq{dSmn}. For large
spins, the spin-orbit interaction can even lead to chaotic motion out
of the orbital plane \cite{Suzuki:1996gm}. This chaotic regime may
significantly affect the power and shape of gravitational radiation
from colliding black holes. It would be interesting to explore similar
effects for the electromagnetic interaction of spinning particles. 

\section*{Acknowledgement}
This research was supported by Natural Sciences
and Engineering Research Canada (NSERC).

\appendix

\section{Spinning ring  in the Lab frame}\label{AppendixA}

\begin{figure}[tbph]
\centering
\includegraphics[scale=0.7]{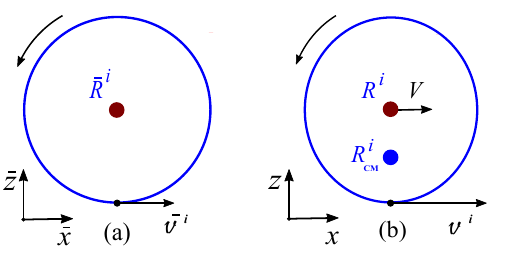}
  \caption{Panel (a) represents the rotating ring in its CM frame. In
    the Lab frame (b) it moves with the velocity $V$. Speeds and
    thus energies of the upper parts are smaller than of
    the lower ones, so the CI shifts below the center of the ring.
\label{TP-Figure-8}}
\end{figure}

As an illustration of physics behind the relativistic shift of the CI
relative to the CM we consider a simple model of a gyroscope as a
rotating ring proposed by Muller \cite{Muller:1991}, 
shown in Fig.~\ref{TP-Figure-8}. 
Let $\rho$ be the radius of the ring and $\mu$ the linear mass
density. In the center of mass frame every element of the ring,
characterized by angle $0\le\psi<2\pi$, moves around a circle in the
$x-z$ plane, 
\be\label{barx}
[\bar{x}(\bar{t}), \bar{y}(\bar{t}), \bar{z}(\bar{t})]
=\rho[\sin(\Omega \bar{t}+\psi), 0, -\cos(\Omega \bar{t}+\psi)],
\ee
with the four-velocity
\be\begin{split}\label{baruv}
&\bar{u}^{\mu}=\bar{\gamma}~ [1, \bar{v}\cos(\Omega \bar{t}+\psi),0, \bar{v}\sin(\Omega \bar{t}+\psi)],\\
&\bar{v}=\Omega\rho  \hh
\bar{\gamma}={1\over\sqrt{1-\bar{v}^2}}.
\end{split}\ee
Here $\bar{x}^{\alpha}$ are coordinates in the  CM frame, $\Omega$ is
the angular velocity of the ring, and its spin tensor is
\be
\bar{S}^{\mu\nu}=\int_0^{2\pi} (\mu\rho d\psi) [\bar{x}^{\mu}\bar{u}^{\nu}-\bar{x}^{\nu}\bar{u}^{\mu}]
\ee
With $
\int_0^{2\pi} d\psi\, \sin,\cos(\Omega \bar{t}+\psi)=0$,
we obtain
\be\begin{split}\label{Sxz}
&\bar{S}^{xz}=2\pi \mu \bar{\gamma} \rho^2 \bar{v}=\bar{E}\Omega\rho^2\equiv\bar{S},
\\
&\bar{S}^{0x}=\bar{S}^{0y}=\bar{S}^{0z}=\bar{S}^{xy}=\bar{S}^{yz}=0.
\end{split}\ee
Invariant mass $M=\sqrt{-P^{\alpha}P_{\alpha}}$ of the ring is
\be\label{barEM}
M=\bar{E}=2\pi\mu\rho\bar{\gamma}.
\ee
Consider this rotating ring in the Lab frame. Lorentz transformation
to the Lab coordinates $x^{\alpha}$ is (assuming that CM moves with
velocity $V$ along $x$)
\be\begin{split}\label{t}
&t=\gamma_\ins{V} (\bar{t}+V\bar{x}) \hh
\gamma_\ins{V}={1/ \sqrt{1-V^2}},\\
&x=\gamma_\ins{V} (\bar{x}+V\bar{t}) \hh y=\bar{y}\hh
z=\bar{z}.
\end{split}\ee
Velocity $v^{i}=dx^i/dt$ of a ring element transforms as
\be
\label{vxyz}
v^x={\bar{v}^x+V\over 1+V\bar{v}^x}\hh 
v^{y,z}={\bar{v}^{y,z} \over \gamma_\ins{V} (1+V\bar{v}^x)}.
\ee
The Lab four-velocity of a ring element  reads
\be
\label{u0xyz}
u^{\mu}={\gamma}~ [1, v^x,v^y,v^z]\hh \gamma=1/\sqrt{1-\bm v^2}.
\ee
\eq{vxyz} and \eq{baruv} give
$\gamma=\bar{\gamma}\gamma_\ins{V}\,(1+V\bar{v}^x)$.
Therefore
\be
\label{utxyz}
u^{\mu} = \bar{\gamma}[\gamma_\ins{V}\,(1+V\bar{v}^x),
\gamma_\ins{V}\,(\bar{v}^x+V), \bar{v}^y, \bar{v}^z ].
\ee
Now compute the component $S^{\alpha 0}$ \eq{Sa0} of the spin tensor,
defining the CM shift. The stress-energy tensor is localized
on the ring so the spatial integration reduces to
the integration over the angle $\psi$. We obtain
\be\label{Si0Tb}
S^{\alpha 0}=\int\limits_0^{2\pi} (\mu\rho d\psi) (x^{\alpha }-R^{\alpha })u^{0}.
\ee
Substitution of \eq{utxyz} and \eq{baruv} leads to $S^{y 0}=0$ and
\be\begin{split}
&S^{x 0}=\mu\rho^2\bar{\gamma}\gamma_\ins{V}^2\int\limits_0^{2\pi}d\psi\,
\sin(\Omega \bar{t}+\psi)[1+V\bar{v}\cos(\Omega \bar{t}+\psi)]
,\\
&S^{z 0}=-\mu\rho^2\bar{\gamma}\gamma_\ins{V}\int\limits_0^{2\pi}d\psi\,
\cos(\Omega \bar{t}+\psi)[1+V\bar{v}\cos(\Omega \bar{t}+\psi)].
\end{split}\ee
All quantities should be evaluated in the Lab frame, at constant $t$ rather than $\bar{t}$.
Following Muller \cite{Muller:1991}, we express them in terms of the Lab time  $t$. This is important
because the CM shift results from two equal \cite{Muller:1991}
contributions: i) Faster moving parts of the ring are heavier; ii)
Time delay depends on the
part of the ring as seen in the Lab. Using
$t=\gamma_\ins{V} (\bar{t}+V\bar{x}),$
\be
\label{cossin}
\cos,\sin (\Omega \bar{t}+\psi)=\cos,\sin\left({\Omega\over\gamma_\ins{v}} t+\psi-V\Omega\bar{x}\right).
\ee
We use $\psi + \Omega t /\gamma_\ins{v} \equiv \varphi$ as the
coordinate on the ring; at $t=\const$ it differs from $\psi$ by a
constant. 
Using \eq{barx} we substitute $\bar{x}$ to \eq{cossin}.
Assuming slow rotation, $\Omega\rho= \bar{v}\ll 1$, we
expand  to the linear order in $\bar{v}$,
\begin{align}
\cos(\Omega \bar{t}+\psi)
&=\cos\varphi+V\bar{v}\,\sin^2\varphi+O(\bar{v}^2),\\
\sin(\Omega \bar{t}+\psi)
&=\sin\varphi-V\bar{v}\,\sin\varphi\cos\varphi+O(\bar{v}^2).
\end{align}
Integrations over $\psi$ and $\varphi$ are  equivalent on
$[0,2\pi)$. We find that $S^{x 0}\sim \int_0^{2\pi}d\varphi\,\sin\varphi = 0$
and
\begin{align}
S^{z 0}&\simeq -\mu\rho^2\bar{\gamma}\gamma_\ins{V}\int\limits_0^{2\pi}d\varphi\,
(V\bar{v} +\cos\varphi)
\nonumber\\
&=-2\pi\mu\rho^2\bar{\gamma}\gamma_\ins{V}V\bar{v}
=-\gamma_V V\bar{S}.
\end{align}
Similarly we calculate the energy,
\be\label{P0a}\begin{split}
P^{0}&=\mu\rho \int\limits_0^{2\pi} d\psi\, u^{0}= \mu\rho \int\limits_0^{2\pi} d\psi\,\gamma_\ins{V}(\bar{u}^0+V\bar{u}^x)\\
&\simeq \mu\rho \bar{\gamma}\gamma_\ins{V}\int\limits_0^{2\pi} d\psi\,(1+V\bar{v}\cos\varphi)=2\pi \mu\rho \bar{\gamma}\gamma_\ins{V}=\gamma_V M.
\end{split}\ee
Here $\bar{S}$ and $M$ are given by \eq{Sxz} and \eq{barEM}.
The CI shift \eq{DeltaR}
of the rotating ring becomes
\be\label{DeltaX5}\Delta R^{x,y,0}=0 \hh 
\Delta R^{z}=-V{\bar{S}\over M}.
\ee
We emphasize that this relativistic derivation holds for all speeds
$-1<V< 1$; only $\bar{v}$ is assumed to be 
small. Eq.~\eq{DeltaX5}  is the relativistic generalization of
Muller's \cite{Muller:1991} and R\k{e}bilas's
\cite{Rebilas:2015}  results. Because $\Delta R^{z}$ in
\eq{DeltaX5} is linear in the spin, it does not depend on the
distribution of the rotating matter. It applies
to any axisymmetric gyroscope, not only a ring.

\section{Motion of the center of mass}\label{AppendixB}

Here we focus on the influence of the classical spin on the orbit of
CM.  
Treating spin as a perturbation, we characterize
the motion of the CM in the direction perpendicular to the zeroth
order orbital plane.
We neglect radiation
effects, proportional to the square of the particle's charge, and
the magnetic moment. The linear momentum obeys  \cite{Teitelboim:1979px}
\begin{align}\label{Frenkel}
{d\over d\tau}\Big(MU^{\alpha}+S^{\alpha\beta}w_{\beta}\Big)&=f^{\alpha},
\\
f^{\alpha}&=qF^{\alpha\beta}U_{\beta}. \label{fa}
\end{align}
The spin affects the motion via the second term in
\eq{Frenkel}. Of higher order in derivatives, it may cause
unphysical instabilities similar to
self-acceleration due to radiation reaction.
To avoid spurious solutions, we treat higher derivatives
perturbatively. We consider the spin to
orbital angular momentum ratio as small, move the
spin-dependent term to the right hand side, and
interpret it as an extra force $\Delta f$,
\be
\Delta f^{\alpha}=-{d\over d\tau}\Big(S^{\alpha\beta}w_{\beta}\Big)
=-w_{\beta}{d \over d\tau}S^{\alpha\beta}
-S^{\alpha\beta}{d \over d\tau}w_{\beta}.
\ee
Because of \eq{dSmn},  antisymmetry of $S^{\alpha\beta}$, and the orthogonality property
$
w^{\alpha}U_{\alpha}=0,
$
the first term vanishes.
For an electromagnetic interaction \eq{fa}, Eq.~\eq{Frenkel} becomes
\be\label{FrenkelA}
M w^{\alpha} = q F^{\alpha\beta}U_{\beta}-S^{\alpha\beta}{d \over d\tau}w_{\beta}
\ee
We look for solutions in the form of a series
\be
w^{\alpha}=w^{\alpha}_{(0)}+w^{\alpha}_{(1)}+\dots,
\ee
and similarly for $R$ and $S$. Note that our goal is to find the
effect of the spin; we compare the motion with that of a spinless
particle of the same energy. Thus we assume that the 4-velocity $U$ is
not perturbed.
In the first two orders, Eqs.~(\ref{Frenkel},\ref{FrenkelA}) reduce to
\ba\label{w0}
M w^{\alpha}_{(0)} &=& q F^{\alpha\beta}U_{\beta},
\\
M w^{\alpha}_{(1)} &=& - S^{\alpha\beta}_{(0)}{d \over d\tau}w_{(0)}{}_{\beta}. \label{w1}
\ea
The zeroth order \eq{w0} is the same as for a spinless
particle, with circular orbits in static external Maxwell fields.
To compare orbits of particles with and without spin,
we assume that the kinetic energy in the Lab frame and the external
field are the same  in both cases. The velocity $V$
and the $\gamma$-factor are therefore treated as fixed.

The first order equation \eq{w1} takes the form
\begin{equation}\label{w1a}
Mw^\alpha_{(1)} =  - {q\over M}S^{\alpha\beta}_{(0)}F_{\beta\lambda}w_{(0)}^\lambda.
\end{equation}
The only non-zero components of the nucleus' field are $F^{i0}$.
Eq.~\eq{DeltaR}  relates $S^{\alpha 0}$ to the CI shift
$\Delta R ^{\alpha}$,
\begin{align}
M w^{i}_{(0)} &= qU^0F^{0i},
\\
-{q\over M}S^{\alpha0}_{(0)}F_{0i}w_{(0)}^{i} 
&= {1\over U^0}S^{\alpha 0}_{(0)}w_{(0)}^2 
= M w_{(0)}^2\Delta R_{(0)}^{\alpha}.
\label{DeltaF}
\end{align}
Equations (\ref{w1},\ref{w1a}) become
\be\label{w1aa}
w^{\alpha}_{(1)} = w_{(0)}^2\Delta R_{(0)}^{\alpha}.
\ee
Thus the first order correction to the CM orbit is described by
\eq{w1a}.   Because  the external field has vanishing $z$ component $F^{z\beta}=0$ the dynamics in $z$ direction decouples from that in the orbital plane and with the same accuracy reduces to
$w^{z}_{(1)} = w^2\Delta R^{z}$.
Substituting
\be
\Delta R^{z}(t)={\tilde{S} V\over M} \sin\gamma\omega t
\ee
from \eq{DeltaR1},  with $w^2=\gamma^4 \omega^4 r^2$, 
leads to
\be
{d^2 z\over dt^2}=\gamma^2\omega^4 r^2{\tilde{S}V\over M} \sin\gamma\omega t.
\ee
CM deviates from the zeroth-order orbital plane by
\be
z=-\omega^2 r^2\,{\tilde{S}V\over M}\sin\gamma\omega t=-V^2\Delta R^{z}(t).
\ee


\end{document}